\documentstyle[11pt,newpasp,twoside,epsf]{article}
\def\edcomment#1{\iffalse\marginpar{\raggedright\sl#1\/}\else\relax\fi}
\reversemarginpar

\begin{document}
\title{The [O {\sc iii}]$\lambda$4363 Emitting Region Obscured by Dusty Tori}
\author{Tohru NAGAO, Takashi MURAYAMA, and Yoshiaki TANIGUCHI}
\affil{Astronomical Institute, Graduate School of Science, Tohoku University,
       Aramaki, Aoba, Sendai 980-8578, Japan}

\begin{abstract}
The emission-line flux ratio of 
[O {\sc iii}]$\lambda$4363/[O {\sc iii}]$\lambda$5007 is a useful diagnostic
for the ionization mechanism and physical properties of narrow-line regions
in active galactic nuclei. However, it is known that simple
photoionization models underpredict this ratio. 
In this contribution, we report on some pieces of evidence that
a large fraction of the [O {\sc iii}]$\lambda$4363 emission arises from
the dense gas, which can be obscured by dusty tori.
Taking this dense component into account, we show that the flux ratio of
[O {\sc iii}]$\lambda$4363/[O {\sc iii}]$\lambda$5007 can be explained
by two-component photoionization models.
\end{abstract}

\section{Introduction}

It has often been considered that the narrow-line regions (NLRs) in 
active galactic nuclei (AGNs) are photoionized by the radiation from
central engines. However, this photoionization scenario has sometimes been 
conflicted with several serious problems.
One of such problems is that any single-zone photoionization models 
underpredict the [O {\sc iii}]$\lambda$4363/[O {\sc iii}]$\lambda$5007 
flux ratio, $R_{\rm O {\sc iii}}$ (e.g., Simpson et al. 1996).
To solve this problem, some models have been proposed; e.g.,
two component models with high-density gas clouds (e.g., Filippenko 1985) or
those with shock-heated regions (e.g., Dopita \& Sutherland 1995).
In addition, some previous studies reported that 
type 1 and type 1.5 Seyfert nuclei (S1s and S1.5s, respectively) tend to
have higher $R_{\rm O {\sc iii}}$ values than type 2 
Seyfert nuclei (S2s) (e.g., Heckman \& Balick 1979).

In order to know where and how the [O {\sc iii}]$\lambda$4363 emission
is radiated, and in order to solve the underprediction problem of
$R_{\rm O {\sc iii}}$, it seems critically important to investigate
why S1s show higher $R_{\rm O {\sc iii}}$ than S2s. 
Therefore, we examine how
the observed value of $R_{\rm O {\sc iii}}$ depends on the AGN type based on
a large sample of AGNs compiled from the literature. Then, we compare
$R_{\rm O {\sc iii}}$ with various parameters to know the nature of the 
[O {\sc iii}]$\lambda$4363 emitting regions in AGNs
(see Nagao, Murayama, \& Taniguchi 2001 for more details).

\section{Results}

\begin{figure}
\plotone{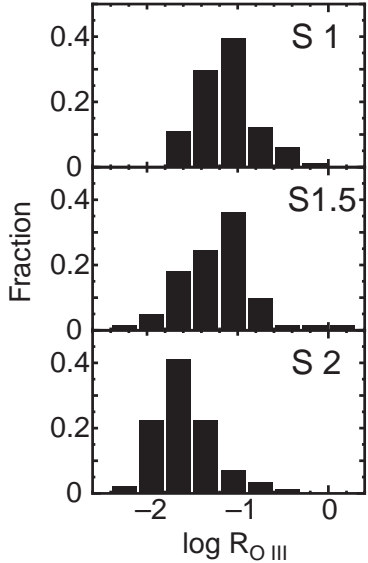}
\caption{Frequency distributions of $R_{\rm O {\sc iii}}$ for the S1s,
         the S1.5s, and the S2s.}
\end{figure}

In order to investigate statistical properties of $R_{\rm O {\sc iii}}$,
we compiled $R_{\rm O {\sc iii}}$ of 82 S1s, 54 S1.5s, and 78 S2s, from 
the literature. The histograms of $R_{\rm O {\sc iii}}$ for each Seyfert
type are shown in Figure 1. It is clearly shown that the S2s
exhibit lower $R_{\rm O {\sc iii}}$ than the S1s and the S1.5s. 
According to the Kolmogorov-Smirnov statistical test, the probability
which the distribution of $R_{\rm O {\sc iii}}$ of S2s and that of
S1s or S1.5s come from the same underlying population is less than
10$^{-6}$.

To explore the origin of this AGN-type dependence of $R_{\rm O {\sc iii}}$
and the nature of the [O {\sc iii}]$\lambda$4363 emitting regions,
we compare $R_{\rm O {\sc iii}}$ with various parameters.
The results are as follows:
  \begin{itemize}
    \item {\it The higher-$R_{\rm O {\sc iii}}$ objects show the hotter 
          mid-infrared (MIR) colors.}
          The hotter MIR colors are thought to be attributed 
          to the hotter dusty grains located at the inner surface of 
          the dusty tori, which can be seen if we see the torus from a
          favored viewing angle (i.e., a nearly face-on view).
          Therefore, this means that the higher-$R_{\rm O III}$ objects are
          seen from a more face-on view toward dusty tori than the 
          lower-$R_{\rm O {\sc iii}}$ objects.
    \item {\it The higher-$R_{\rm O {\sc iii}}$ objects show the stronger 
          {\rm [Fe {\sc vii}]$\lambda$6087 and [Fe {\sc x}]$\lambda$6374}
          emission.} 
          Since a large fraction of high-ionization emission lines is 
          thought to arise from dense gas clouds obscured by the dusty tori 
          (e.g., Murayama \& Taniguchi 1998a), 
          the higher-$R_{\rm O {\sc iii}}$ can be 
          attributed to the significant flux contribution from such a 
          high-density gas cloud.
    \item {\it The higher-$R_{\rm O {\sc iii}}$ objects tend to show the
          larger {\rm FWHM([O {\sc iii}]$\lambda$4363)/
          FWHM([O {\sc iii}]$\lambda$5007)} ratios.}
          This also suggests that the [O {\sc iii}]$\lambda$4363 emitting 
          regions are located at inner regions compared to the 
          [O {\sc iii}]$\lambda$5007 emitting regions.
    \item {\it The S1s have wider {\rm FWHM([O {\sc iii}]$\lambda$4363)} 
          and larger ratio of {\rm 
          FWHM([O {\sc iii}]$\lambda$4363)/FWHM([O {\sc iii}]$\lambda$5007)}
          than the S2s.} This suggests that the [O {\sc iii}]$\lambda$4363
          emitting regions are located at inner regions compared to the 
          [O {\sc iii}]$\lambda$5007 emitting regions and have 
          an anisotropic property.
  \end{itemize}
All of these facts can be consistently understood when we introduce the
high-density gas clouds located close to the nucleus, which emit a large 
fraction of the [O {\sc iii}]$\lambda$4363 emission. Since these clouds
may suffer significantly from the obscuration by dusty tori in S2s, 
we conclude that the AGN-type
dependence of $R_{\rm O {\sc iii}}$ is attributed to the viewing-angle
dependence of the visibility of the high-density gas clouds.

\section{Discussion}

\begin{figure}
\plotone{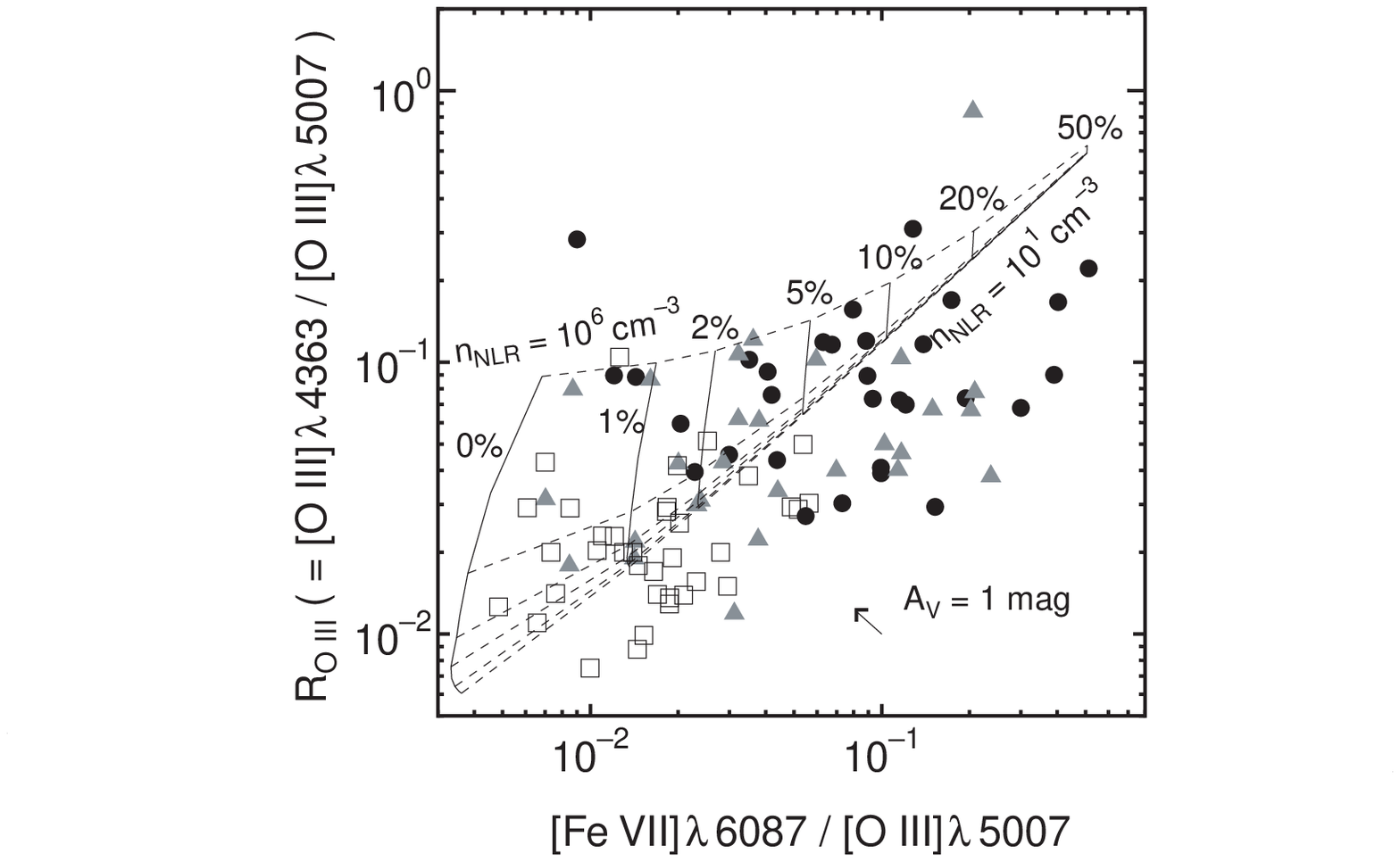}
\caption{Diagram of $R_{\rm O {\sc iii}}$ vs. 
         [Fe {\sc vii}]$\lambda$6087/[O {\sc iii}]$\lambda$5007. The data
         of the S1s, the S1.5s, and the S2s are shown by filled circles,
         gray triangles, and open squares, respectively. 
         Loci of our model calculations are superposed in the figure.
         The fraction of the flux contribution from the dense component into
         the [O {\sc iii}]$\lambda$5007 emission is shown 
         (0\%, 1\%, ..., 50\%). The data points will move on the diagram 
         as shown by the arrow if the extinction correction of $A_V$ = 1.0 mag
         is applied.}
\end{figure}

To examine whether or not the viewing-angle dependence of the visibility
of the high-density gas can account for the observed difference in
$R_{\rm O {\sc iii}}$ between S1s and S2s in the framework of
the photoionization scheme,
we carry out dual-component photoionization model 
calculations following the manner of Murayama \& Taniguchi (1998b).
This method takes account of such high-density gas clouds as a strong
[O {\sc iii}]$\lambda$4363 emitter, in addition to the typical NLR component.

We perform model calculations using the spectral synthesis code
$Cloudy$ version 90.04 (Ferland 1996). We assume uniform density gas 
clouds with a plane-parallel geometry. The dense component (DC) is 
assumed to be truncated clouds to avoid unusually strong 
[O {\sc i}] emission. Here we assume $n_{\rm DC} = 10^7$ cm$^{-3}$.
We perform several model runs covering 
$10^1$ cm$^{-3} \leq n_{\rm NLR} \leq 10^6$ cm$^{-3}$.
The ionization parameter of the NLR component is assumed as $U_{\rm NLR}$
= 10$^{-2}$. The ionization parameter and the hydrogen column density
of DC are determined by the following two conditions:
([Fe {\sc x}]/[Fe {\sc vii}])$_{\rm DC}$ = 0.8 and
([Fe {\sc vii}]/[O {\sc iii}]$\lambda$5007)$_{\rm DC}$ = 1.0.
The former ratio is the typical value of Seyfert galaxies
(Nagao, Taniguchi, \& Murayama 2000), and the latter
condition is introduced by Murayama \& Taniguchi (1998b).
As a result, $U_{\rm DC} = 10^{-1.48}$ and $N_{\rm DC} = 10^{20.76}$ cm$^{-2}$
are adopted. The calculations are stopped when the gas temperature falls to 
4000K for the NLR component. We set the gas-phase element abundances
to be solar ones. We adopt the power-law continuum as $\alpha$ = --1.5
between 10 $\mu$m and 50 keV in the form $f_{\nu} \propto \nu^{\alpha}$.
The spectral index is set to $\alpha$ = 2.5 at lower energy 
(i.e., $\lambda \geq 10 \mu$m) and to $\alpha$ = --2 at higher energy 
(i.e., $h\nu \geq$ 50 keV). The fraction of DC to the NLR component is 
treated as a free parameter in our calculations.

We present our model calculations and compare them with the observations
in a diagram of $R_{\rm O {\sc iii}}$ versus
[Fe {\sc vii}]/[O {\sc iii}]$\lambda$5007 (Figure 2). 
We find that the model grids 
are roughly consistent with the observations if we take the effects of 
the correction for the extinction into account. Though the dispersion of 
observation appears to be larger than the model grids, this may be
attributed to the fact that the parameters, e.g., $U_{\rm NLR}$ and 
$n_{\rm DC}$, are different from object to object. It is shown that the
$R_{\rm O {\sc iii}}$ of the S1s can be explained by introducing a 
5\% -- 20\% contribution from DC while the $R_{\rm O {\sc iii}}$ of the S2s
can be explained by introducing a 0\% -- 2\% contribution from DC.

Through the dual-component photoionization model calculations,
we conclude that (1) the origin of the AGN-type dependence of
$R_{\rm O {\sc iii}}$ is the viewing-angle dependence (i.e., the 
AGN-type dependence) of the visibility of the dense gas clouds which can
be hidden by dusty tori, and (2) the underprediction problem of
$R_{\rm O {\sc iii}}$ is solved taking such high-density gas clouds into
account.

\end{document}